\documentclass{llncs}
\usepackage{mathptmx}       
\usepackage{helvet}         
\usepackage{courier}        
\usepackage{type1cm}        
\usepackage{makeidx}         
\usepackage{graphicx}        
\usepackage{multicol}        
\usepackage[bottom]{footmisc}
\usepackage{subfigure}
\usepackage{amsfonts}
\usepackage[cmex10]{amsmath}
\usepackage{float}
\begin{document}

\title{Max-Plus Opinion Dynamics With Temporal Confidence}
\titlerunning{Max-Plus Opinion Dynamics}  
%
\author{Daniel Feinstein\inst{1} \and Ebrahim Patel \inst{1}}
\authorrunning{Daniel Feinstein et al.} 
%
\tocauthor{Ebrahim Patel}
\institute{University of Oxford, Andrew Wiles Building, Woodstock Rd, Oxford OX2 6GG\\
	\email{danielfeinstein12@gmail.com}}
	
\maketitle    
\section{Introduction}
Often in the setting of human-based interactions, the existence of a temporal hierarchy of information plays an important role in diffusion and opinion dynamics within communities \cite{anchoring}. For example at the individual agent level, more recently acquired information may exert greater influence during decision-making processes \cite{recency}. To facilitate further exploration of this effect, we introduce an efficient method for modelling temporally asynchronous opinion updates, where the timing of updates depends on the timing of incoming opinion states received from neighbours. The framework enables the introduction of \textit{information arrival-time lag} by means of \textit{lag-vectors}. These are used to weight the relevance of information received by agents, based on the delay between its receipt and the subsequent opinion update. The temporal dynamics (i.e. the times at which information is transmitted) are governed by an underlying algebraic structure called max-plus algebra (\cite{maxplus}, \cite{maxplus2}). We investigate the resulting continuous opinion dynamics under the max-plus regime using a modified Hegselmann-Krause model \cite{HK}, replacing the conventional confidence-interval based on the distance between opinions with one based instead on the recency of information received by agents.
\\
\\
Our model works as follows: at time-step $k = 0$, each agent (represented by a node in a network) is assigned an initial opinion from the interval $[0,1]$ uniformly at random and transmits this value to all neighbours in the network. If an edge from agent $i$ to $j$ exists, the information leaving agent $i$ arrives at $j$ after $A_{ji}$ time units (e.g. minutes), where $A$ is the max-plus adjacency matrix (which is nothing more than the transpose of the conventional adjacency matrix). After sending their current opinion, each agent enters a dormant period where it waits to receive all incoming opinion values. Once received, agents update their opinions before immediately re-sending their new values to all neighbours (possibly at different times), and this process continues for a desired duration. 
\\
\\
The event-times (the times at which opinion updates occur) can be conveniently modelled using the max-plus algebra which we denote ${R}_\infty$ . Let $\vec{x}(k) \in {R}_\infty^{(n\times 1)}$ denote the vector of the $(k+1)_{st}$ time agents communicate their opinions. Then $\vec{x}_i(k)$ is the time of the $(k+1)_{st}$ transmission of agent $i$ and is defined, in line with the above description, by:
\begin{equation}
\vec{x}_i(k) = \max \{\vec{x}_j(k-1) + A_{ij} : j = 1,\ldots n \}, \text{ for all } k\geq 1.
\end{equation}
In the notation of the max-plus algebra this becomes,
\begin{equation}\vec{x}(k) = A\otimes \vec{x}(k-1), \text{ for all } k\geq 1.
\end{equation}
We also conveniently model the number of time units agent $i$ has been sitting on the opinion value received from agent $j$ before its next update:
 \begin{equation}
 \vec{\xi}(k,i) = \vec{x}_i(k)\vec{I} -\vec{x}(k-1) - A_i^{T}
 \end{equation}
 where $\vec{I}$ is the $(n\times 1)$ unit column vector and $A_i^T$ is the transposed $i_{th}$ row of $A$.
We refer to the vector above as the \textit{lag-vector} for opinions arriving at agent $i$, having been sent neighbours at time-step $(k-1)$.
\\
\\
To simulate the resulting opinion dynamics, we modify the Hegselmann-Krause model \cite{HK} (which from here, we refer to just as the HK model) to incorporate the lag-vectors. At each time-step $(k+1)$, every agent $i$ updates their opinion according the following update-rule: \begin{equation}\label{a}
o_i(k+1) = \big|\mathcal{N}\big(i, k\big)\big|^{-1} \sum_{j\in\mathcal{N}\big(i, k\big)} o_j(k),
\end{equation}
where $\mathcal{N}(i,k) = \{1\leq j \leq n \big | 0\leq \vec{\xi}_j(i,k)\leq \epsilon  \}$, i.e. the set of $i's$ neighbours whose opinion values are sat on by $i$ for at most $\epsilon$ time units. Note the standard, unmodified HK model update-rule is given by replacing $\mathcal{N}(i,k)$ with ${\mathcal{M}}\big(i, k\big) = \{1\leq j\leq n \big| |o_i(k)-o_j(k)| \leq \epsilon\}$.
\\
\\
To summarize the entire process for each time-step: each agent waits until it has received all incoming information (modelled by equation 2). On receipt of the final incoming opinion value for the current time-step, agents update their opinion using the modified HK update-rule (equation 4) and send this to all neighbours. 
\section{Results}
We show via extensive computational simulations that the updated HK model (using the temporally bounded confidence-interval in equation 4) supports multi-opinion consensus clusters despite the absence of the conventional confidence-interval based on the distance between neighbouring opinions (Fig 1). This is significant because it demonstrates that opinion fragmentation is possible even with seemingly innocent sorting of content based only on recency considerations.
Simulations are carried out on random weighted strongly-connected and directed Barabási–Albert, Erdős–Rényi and Watts-Strogatz graphs consisting in each case of 100 nodes. Furthermore, we examine typical behaviours emerging from varying the threshold beyond which agents fail to take opinions from their neighbours into account, i.e. from varying $\epsilon$, while keeping all other initial conditions fixed. Two noticeably distinct regimes emerge. The first is a gradual transition from multiple consensus clusters to a single global consensus. As epsilon is increased, the number of consensus clusters grows slowly, until $\epsilon$ becomes large enough for a single global consensus to form. The second regime is of a more abrupt and discrete change. The number of consensus clusters remains fixed before a sudden transition occurs beyond some critical value of $\epsilon<1$, where multiple opinion clusters suddenly collapse into global consensus. 

\begin{figure}[H]
\centering
\parbox{5cm}{
\includegraphics[width=5cm]{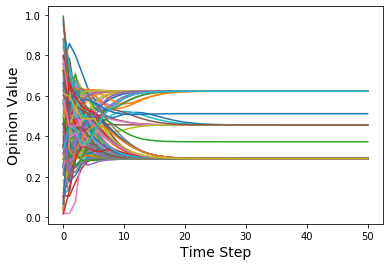}
\caption{Multi-opinion consensus clusters emerging using the modified HK update-rule with $\epsilon = 1$. The simulation was carried on a random weighted strongly-connected directed Barabási–Albert graph with 100 nodes, Barabási–Albert parameter of 2, and edge-weights drawn uniformly at random from $\{1,...,20\}$.}

\label{fig:2figsA}}
\qquad
\begin{minipage}{5cm}
\includegraphics[width=5cm]{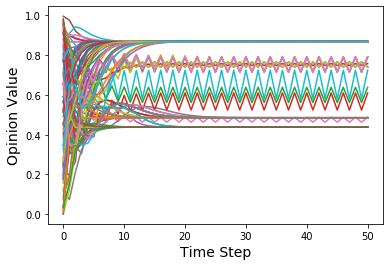}
\caption{Oscillatory behaviour of opinions emerging using the modified HK update-rule with $\epsilon = 1$. The simulation was carried on a random weighted strongly-connected directed Barabási–Albert graph with 100 nodes, Barabási–Albert parameter of 2, and edge-weights drawn uniformly at random from $\{1,...,20\}$.}
\label{fig:2figsB}
\end{minipage}
\end{figure}

We also uncover a new phenomenon arising from the dynamics using the modified HK update-rule (equation 4), whereby multi-opinion consensus clusters emerge alongside groups of agents exhibiting opinion values which oscillate in time with a regular period (Fig 2). This type of behaviour is not supported by the standard HK model under any circumstance. Using a max-plus periodicity result, we explain this phenomenon analytically by showing that lag-vectors are in fact periodic, with the period being dependent on circuits within the network. Namely, we prove the following:
if $A$ is a strongly-connected max-plus adjacency matrix, there exist positive constants $k_c$ and $C$ such that $\vec{\xi}(k+C, i) = \vec{\xi}(k, i)$ for all $k_c\leq k$ and agents $i$.
This provides analytical insight to characterise neighbourhood structures within the network which are susceptible to experiencing periodicity of
opinion states.


\begin{thebibliography}{10}
\providecommand{\url}[1]{\texttt{#1}}
\providecommand{\urlprefix}{URL }
\bibitem{anchoring}
Hartmann, Stephan, and Soroush Rafiee Rad. `Anchoring in deliberations.' Erkenntnis (2019): 1-29.

\bibitem{recency}
Zdep, Stanley, and Warner Wilson. `Recency effects in opinion formation.' Psychological reports 23.1 (1968): 195-200.

\bibitem{maxplus}
Hogben, Leslie, ed. Handbook of linear algebra. CRC press, 2006.

\bibitem{maxplus2}
Heidergott, Bernd, Geert Jan Olsder, and Jacob Van Der Woude. Max Plus at work: modeling and analysis of synchronized systems: a course on Max-Plus algebra and its applications. Vol. 48. Princeton University Press, 2014.

\bibitem{HK}
Hegselmann, Rainer, and Ulrich Krause. `Opinion dynamics and bounded confidence models, analysis, and simulation.' Journal of artificial societies and social simulation 5.3 (2002).



\end{thebibliography}
\end{document}